\documentclass[onecolumn,showpacs]{revtex4}

\usepackage{graphicx}% Include figure files
\usepackage{dcolumn}% Align table columns on decimal point
\usepackage{bm}% bold math

\input epsf

\begin{document}

\title{\Large Naked Singularities in Higher Dimensional Szekeres
space-time}

\author{\bf Ujjal Debnath}
\email{ujjaldebnath@yahoo.com}
\author{\bf Subenoy Chakraborty}
\email{subenoyc@yahoo.co.in}

 \affiliation{Department of
Mathematics, Jadavpur University, Calcutta-32, India.}

\date{\today}

\begin{abstract}
In this paper we study the quasi-spherical gravitational collapse
of ($n+2$)-dimensional Szekeres space-time. The nature of the
central shell focusing singularity so formed is analyzed by
studying both the radial null and time-like geodesic originated
from it. We follow the approach of Barve et al to analyze the null
geodesic and find naked singularity in different situations.
\end{abstract}

\pacs{0420D, 0420J, 0470B}

\maketitle

\section{{\protect\normalsize \textbf{Introduction }}}
In recent years, there is an exhaustive study of gravitational
collapse of inhomogeneous spherical dust [1-4]. It is mainly
concentrated to central shell focusing singularity. The behaviour
of the singularity (black hole or naked singularity) has been
analyzed mainly by the studying of outgoing radial null geodesic
and the strength is measured using due to Tipler [5]. It has been
shown that appearance of naked singularity strongly depends on
the initial data [6-10]. \\

In contrast, there is very little information about singularity
formation and structure in non-spherical collapse. Here horizon
will be formed if and only if the \textit{hoop conjecture} [11]
is satisfied. Most of the studied so far in non-spherical collapse
are by numerical analysis and for some definite shape of the
gravitating mass [12-16].\\

However, the study of gravitational collapse for quasi-spherical
Szekeres space-time [17] was started by Szekeres himself [18]
long ago. Afterwards, it was studied further by Joshi et al [19]
and extensively by Goncalves [20].  Recently, we have obtained
solutions for ($n+2$)-dimensional Szekeres space-time with
perfect fluid (or dust) as the matter content [21]. Subsequently,
we have also analyzed the gravitational collapse for the above
dust solution to examine the local behaviour of the singularity so
formed [22].\\

In this paper, we extend our study for global characteristic of
the singularity by studying both null and time like geodesic
originated from the singularity using the formalism of Barve et
al [23]. The paper is organized as follows: In section II we have
written the basic equations and regularity conditions. In section
III and IV, we have studied the radial null and time-like
geodesics originated from the singularity. We have studied the
local visibility of the singularity for radial null and time-like
geodesics due to collapse in section V. Finally the paper ends with a short discussion.\\

\section{{\protect\normalsize \textbf{Basic Equations and regularity
conditions}}} The metric ansatz for ($n+2$)D Szekeres space-time
is given by [21]

\begin{equation}
ds^{2}=dt^{2}-e^{2\alpha}dr^{2}-e^{2\beta}\sum^{n}_{i=1}dx_{i}^{2}
\end{equation}

where $\alpha$ and $\beta$ are functions of all the ($n+2$)
space-time co-ordinates with expression [21] (where $\beta'\ne 0$)

\begin{equation}
e^{\alpha}=R'+R~\nu'
\end{equation}

and

\begin{equation}
e^{\beta}=R(t,r)~e^{\nu(r,x_{1},...,x_{n})}
\end{equation}

Here $R$ satisfied the differential equation

\begin{equation}
\dot{R}^{2}=f(r)+\frac{F(r)}{R^{n-1}}
\end{equation}

and the expression for $\nu$ is

\begin{equation}
e^{-\nu}=A(r)\sum_{i=1}^{n}x_{i}^{2}+\sum_{i=1}^{n}B_{i}(r)x_{i}+C(r)
\end{equation}

with the restriction

\begin{equation}
\sum_{i=1}^{n}B_{i}^{2}-4AC=f(r)-1
\end{equation}

for the arbitrary functions $A(r),~ B_{i}(r)$ and $C(r)$. Also in
the expression (4), $f(r)$ and $F(r)$ are arbitrary functions of
$r$ alone. But due to complexity of the problem we shall restrict
ourselves to marginally bound case only (i.e., $f(r)=0$). In this
case equation (4) can easily integrated to give

\begin{equation}
R=\left[r^{\frac{n+1}{2}}-\frac{n+1}{2}\sqrt{F(r)}~t\right]^{\frac{2}{n+1}}
\end{equation}

and the energy density for the matter field

\begin{equation}
\rho(t,r,x_{1},...,x_{n})=\frac{n}{2}~\frac{F'+(n+1)F\nu'}{R^{n}(R'+R\nu')}
\end{equation}

takes the form

\begin{equation}
\rho_{i}(r,x_{1},...,x_{n})=\rho(0,r,x_{1},...,x_{n})=
\frac{n}{2}~\frac{F'+(n+1)F\nu'}{r^{n}(1+r\nu')}
\end{equation}

initially at $t=0$, where we have chosen the scaling of $R$ as

$$
R(0,r)=r.
$$

Now assuming that the collapse starts from a regular initial
hypersurface so we take the following series forms for $F(r)$,
$\rho_{i}(r)$ and $\nu'(r)$ [21]

\begin{equation}
F(r)=\sum_{j=0}^{\infty}F_{j}~r^{n+j+1}~~,
\end{equation}

\begin{equation}
\rho_{i}(r)=\sum_{j=0}^{\infty}\rho_{j}~r^{j}
\end{equation}

and

\begin{equation}
\nu'(r)=\sum_{j=-1}^{\infty}\nu_{j}~r^{j},~~~(\nu_{_{-1}}\ge -1)
\end{equation}

where the coefficients are related as

\begin{eqnarray*}
\rho_{0}=\frac{n(n+1)}{2}F_{0},~~\rho_{1}=\frac{n}{2}\left(n+1+\frac{1}{1+
\nu_{_{-1}}}\right)F_{1},
\end{eqnarray*}
\vspace{-5mm}

\begin{eqnarray*}
\rho_{2}=\frac{n}{2}\left[\left(n+1+\frac{2}{1+
\nu_{_{-1}}}\right)F_{2}-\frac{F_{1}\nu_{_{0}}}{(1+\nu_{_{-1}})^{2}}\right],
\end{eqnarray*}
\vspace{-5mm}

\begin{equation}
\rho_{3}=\frac{n}{2}\left[\left(n+1+\frac{3}{1+
\nu_{_{-1}}}\right)F_{3}-\frac{2F_{2}\nu_{_{0}}}{(1+\nu_{_{-1}})^{2}}-
\frac{(1+\nu_{_{-1}})\nu_{_{1}}-\nu_{_{0}}^{2}}{(1+\nu_{_{-1}})^{3}}F_{1}\right],
\end{equation}

$$...~~...~~...~~...~~...~~...~~...~~...,$$

$$OR$$

\begin{eqnarray*}
\rho_{0}=\frac{n}{2}\left[\frac{F_{1}}{\nu_{0}}+(n+1)F_{0}\right],~~
\rho_{1}=\frac{n}{2}\left[\frac{2F_{2}}{\nu_{0}}+\left\{(n+1)-
\frac{\nu_{1}}{\nu_{0}^{2}}\right\}F_{1}\right],
\end{eqnarray*}
\vspace{-5mm}

\begin{eqnarray*}
\rho_{2}=\frac{n}{2}\left[\frac{3F_{3}}{\nu_{0}}+\left\{(n+1)-
\frac{2\nu_{1}}{\nu_{0}^{2}}\right\}F_{2}+\left(\frac{\nu_{1}^{2}}{\nu_{0}^{3}}-
\frac{\nu_{2}}{\nu_{0}^{2}}\right)F_{1}\right],
\end{eqnarray*}
\vspace{-5mm}

\begin{equation}
\rho_{3}=\frac{n}{2}\left[\frac{4F_{4}}{\nu_{0}}+\left\{(n+1)-
\frac{3\nu_{1}}{\nu_{0}^{2}}\right\}F_{3}+2\left(\frac{\nu_{1}^{2}}{\nu_{0}^{3}}-
\frac{\nu_{2}}{\nu_{0}^{2}}\right)F_{2}+\left(\frac{2\nu_{1}\nu_{2}}{\nu_{0}^{3}}-
\frac{\nu_{3}}{\nu_{0}^{2}}-\frac{\nu_{1}^{3}}{\nu_{0}^{4}}\right)F_{1}\right],
\end{equation}

$$...~~...~~...~~...~~...~~...~~...~~...,$$

according as~~ $\nu_{_{-1}}>-1$~ or ~$\nu_{_{-1}}>=-1$.\\

The singularity curve $t=t_{s}(r)$ for the shell focusing
singularity is characterized by

$$
R(t_{s}(r),r)=0
$$

and we have

\begin{equation}
t_{s}(r)=\frac{2}{n+1}\frac{r^{\frac{n+1}{2}}}{\sqrt{F(r)}}
\end{equation}

where $t_{0}=\frac{2}{(n+1)\sqrt{F_{0}}}$ is the time for the
central shell focusing singularity. However, near $r=0$, above
singularity curve can be approximately written as

\begin{equation}
t_{s}(r)=t_{0}-\frac{F_{m}}{(n+1)F_{0}^{3/2}}~r^{m}
\end{equation}

where $m~(\ge 1)$ is an integer and $F_{m}$ is the first
non-vanishing term beyond $F_{0}$.\\

\section{{\protect\normalsize \textbf{Radial Null Geodesics}}}

The equation of the outgoing radial null geodesic (ORNG) which
passes through the central singularity in the past is taken as
(near $r=0$)

\begin{equation}
t_{ORNG}=t_{0}+a r^{\xi},
\end{equation}

to leading order in $t$-$r$ plane with $a>0,\xi>0$. Now to
visualize the singularity the time $t$ in the geodesic (17) should
be less than $t_{s}(r)$ in equation (16). Hence comparing these
times we have the restrictions

\begin{equation}
\xi\ge m ~~~~ \text{and} ~~~~ a<-\frac{F_{m}}{(n+1)F_{0}^{3/2}}
\end{equation}

Further, for the metric (1) an outgoing radial null geodesic
should satisfy

\begin{equation}
\frac{dt}{dr}=R'+R~\nu'
\end{equation}

We shall now examine the feasibility of the null geodesic
starting from the singularity with the above restrictions for the
following two cases namely, (i) $\xi>m$ and (ii) $\xi=m$.\\

When $\xi>m$  then near $r=0$ the solution for $R$ in (7)
simplifies to

\begin{equation}
R=\left(-\frac{F_{m}}{2F_{0}}\right)^{\frac{2}{n+1}}r^{\frac{2m}{n+1}+1}
\end{equation}

Now combining (17) and (20) in equation (19) we get (upto leading
order in $r$)

\begin{equation}
a~\xi~
r^{\xi-1}=\left(\nu_{_{-1}}+1+\frac{2m}{n+1}\right)\left(-\frac{F_{m}}{2F_{0}}\right)
^{\frac{2}{n+1}}r^{\frac{2m}{n+1}}
\end{equation}

which implies

\begin{equation}
\xi=1+\frac{2m}{n+1}~~~ \text{and}~~~
a=\frac{(\nu_{_{-1}}+1+\frac{2m}{n+1})}{(1+\frac{2m}{n+1})}\left(-\frac{F_{m}}
{2F_{0}}\right)^{2/(n+1)}
\end{equation}

We note that if $\nu_{_{-1}}=0$ then the above results are same
as TBL model [9], so we restrict ourselves to $\nu_{_{-1}}\ne 0$.
As $\xi>m$ so from (22)

$$
m<\frac{n+1}{n-1}
$$

Since $m$ is an integer so we have the following possibilities:\\

$ ~~~~~~~~~~~~~~~~n=2~ (4D \text{~space-time}): ~~ m=1, 2  ~~~, \xi=\frac{4}{3}, \frac{7}{3}$\\
$ ~~~~~~~~~~~~~~~~~~~~n\ge 3~ (5D \text{~or~ higher dim.
space-time}): ~~ m=1 ~~~, \xi=1+\frac{2}{n+1} $.\\

Now for $\nu_{_{-1}}>-1$, if we assume that the initial density
gradient falls off rapidly to zero at the centre then we must have
$\rho_{1}=0, \rho_{2}<0$ i.e., $F_{1}=0, F_{2}<0$. Hence the
least value of $m$ should be 2. Thus for a self consistent
outgoing radial null geodesic starting from the singularity we
must have $n=2$ only. So the singularity will be naked only in
four dimension. However, if we relax the restriction on the
initial density and simply assume that the initial density
gradually decreases as we go away from the centre (i.e.,
$\rho_{1}<0$) then $F_{1}<0$ and $m=1$ is the possible solution.
Therefore, naked singularity is possible in any dimension. However
for $\nu_{_{-1}}=-1$, the assumption about the initial density gradient
($\rho_{1}=0, \rho_{2}<0$) does not implying $F_{1}=0$. Hence here $m=1$
is a possible solution for all dimension with $F_{1}<0$. Therefore naked
singularity may occur in any dimension for this choice of the parameter $\nu_{-1}$.\\

Now we shall study the second case namely $\xi=m$. Here also for
$\nu_{_{-1}}=0$ we have identical results as in TBL model [9]. For
$\nu_{_{-1}}\ne 0$, using the same procedure as above we get
$$
m=\frac{n+1}{n-1}
$$
and

\begin{equation}
a=-\frac{1}{m}\left(-\frac{n+1}{2}\sqrt{F_{0}}~a-\frac{F_{m}}{2F_{0}}
\right)^{\frac{1-n}{1+n}}
\left[(\nu_{_{-1}}+m)\frac{F_{m}}{2F_{0}}+\frac{1}{2}(n+1)(\nu_{_{-1}}+1)\sqrt{F_{0}}~a\right]
\end{equation}

Hence integral solution for $m$ is possible only for $n=2$ and 3.
So consistent outgoing radial null geodesic through the central
shell focusing singularity is possible only upto five dimension
i.e., we cannot have naked singularity for higher dimensional
space-time greater than five. It is to be noted that this conclusion
is not affected by the value of the parameter $\nu_{_{-1}}$.\\

Now we shall examine whether the restriction (18) for $a$ is
consistent with the expression $a$ in equation (23). In fact
equation (23) takes the form

\begin{equation}
2^{\frac{2}{n+1}}bm=-\left[-(n+1)b-\zeta\right]^{-\frac{1}{m}}\left[(\nu_{_{-1}}+m)\zeta+
(n+1)(\nu_{_{-1}}+1)b\right]
\end{equation}

with the transformation

\begin{equation}
a=bF_{0}^{\frac{1}{n-1}},~~~~F_{m}=\zeta F_{0}^{\frac{m}{2}+1}.
\end{equation}

Since equation (24) is a real valued equation of $b$, so we must
have
$$
(n+1)b+\zeta<0~,
$$

which using (25) gives us the restriction on $a$ in equation (18).
Hence the geodesic (17) will have consistent solution for $a$ and
$m$. So the above conclusion regarding the formation of naked
singularity is justified. Further, introducing the variable
$\phi$ by the relation

\begin{equation}
\phi=-(n+1)b-\zeta
\end{equation}

\begin{figure}
\includegraphics[height= 3.9in]{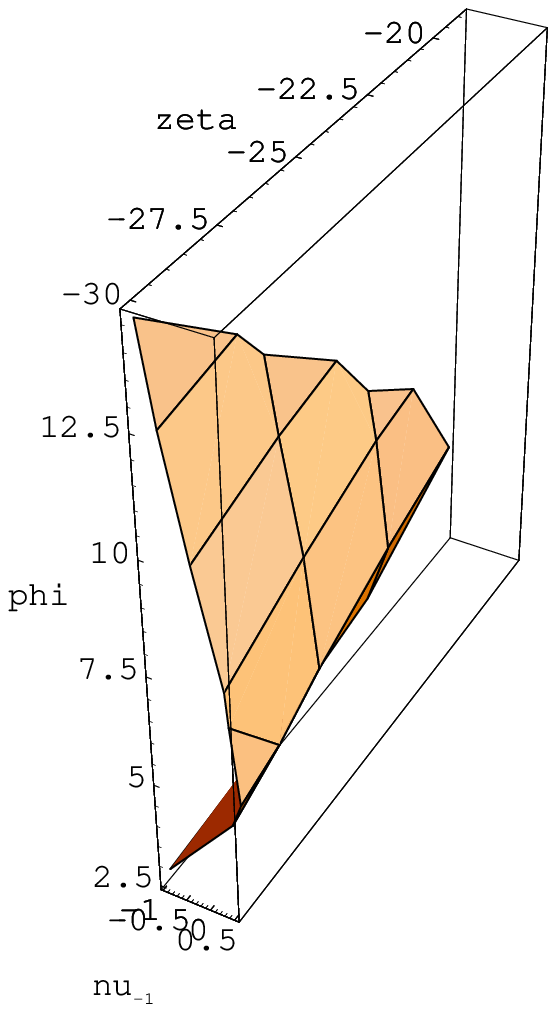}~~~~~~~~~~~~
\includegraphics[height= 3.9in]{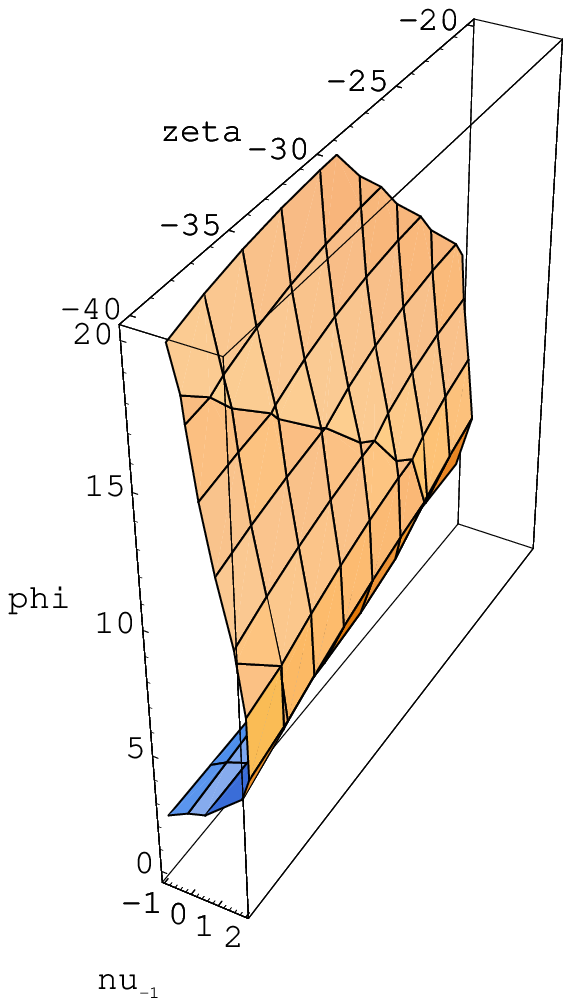}
\\\vspace{5mm}
Fig.1~~~~~~~~~~~~~~~~~~~~~~~~~~~~~~~~~~~~~~~~~~~~~~~~~~Fig.2\\
\vspace{5mm}
\includegraphics[height= 2.5in]{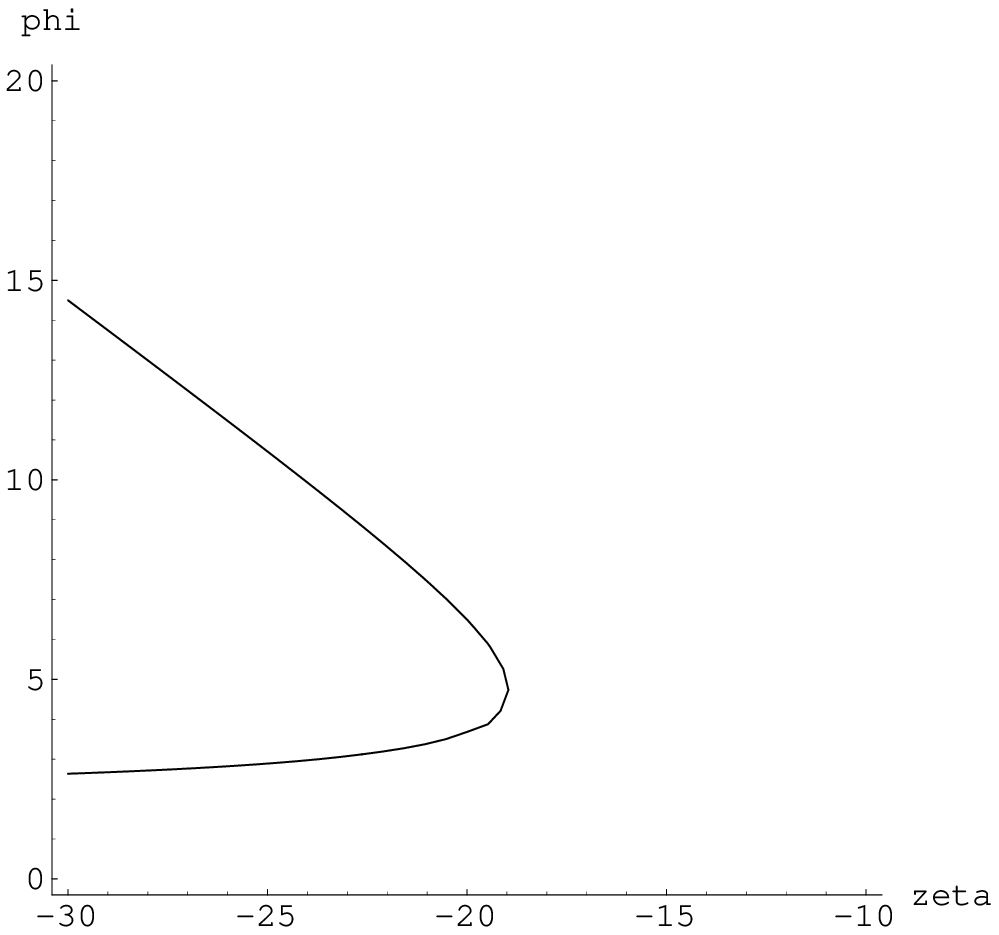}~~~~~~~~~
\includegraphics[height= 2.5in]{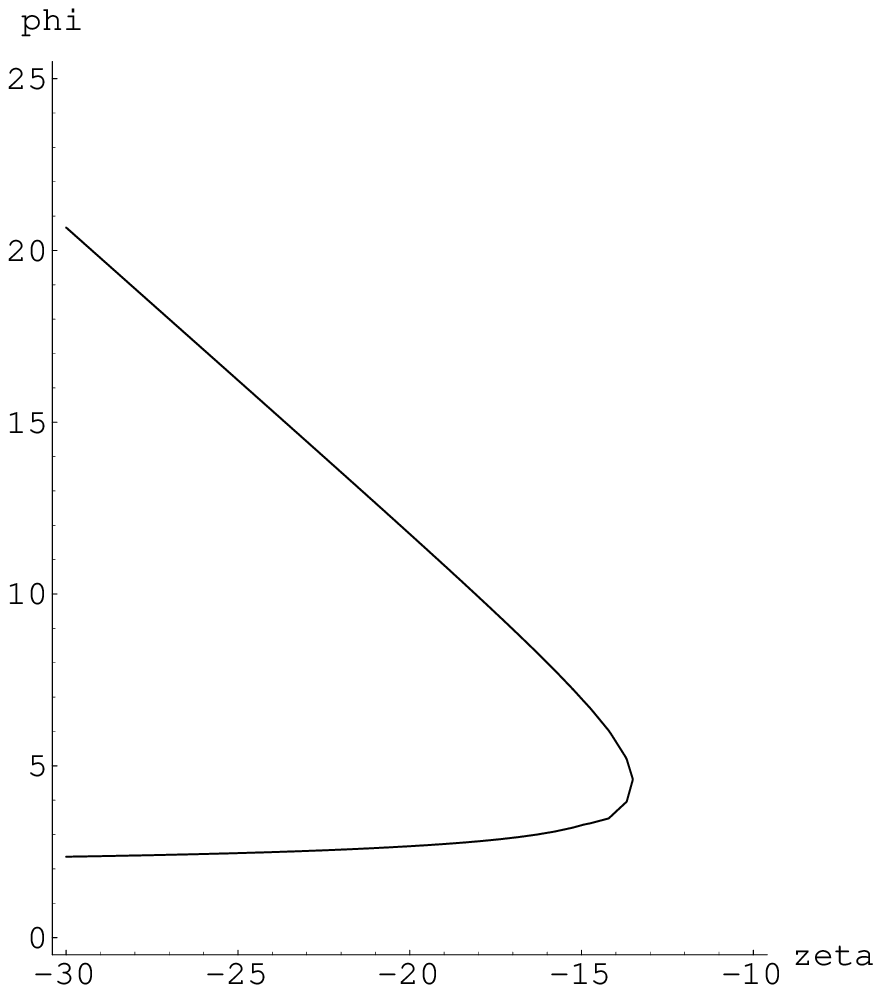}
\\\vspace{5mm}
Fig.3~~~~~~~~~~~~~~~~~~~~~~~~~~~~~~~~~~~~~~~~~~~~~~~~~~Fig.4\\
\vspace{4mm} {In Figs.1 and 2, we have shown the variation of
$\phi~(0<\phi<-\zeta)$ for the variations of $\nu_{_{-1}}(>-1)$
and $\zeta(<0)$ for four and five dimensions (i.e., $n=2$ and 3)
respectively. However for $\nu_{_{-1}}=-1$, the dependence of
$\phi$ over $\zeta$ has been presented in Figs.3 and 4 for $n=2$
and 3 respectively.~~~~~~~}\\
\vspace{1mm}
\end{figure}

We have seen from equation (24)

\begin{equation}
4\phi^{n-1}(\zeta+\phi)^{n+1}=[2\zeta-(n-1)(\nu_{_{-1}}+1)\phi]^{n+1}
\end{equation}

with the restriction ~$0<\phi<-\zeta$.\\

Finally, we investigate whether there is only one null geodesic
emanating from the singularity (i.e., when singularity is visible
for an infinitesimal time) or it is possible to have an entire
family of geodesic through the singularity (this occurs when
singularity is visible for an infinite time). So we write the
equation for the outgoing radial null geodesics to next order as
\begin{equation}
t=t_{0}+a r^{\xi}+h r^{\xi+\sigma}
\end{equation}

where as before $a, \xi, h, \sigma$ are all positive. Then
proceeding in the same way we have\\

(i) when $\xi>m$, $a$ and $\xi$ have the same expressions as in
equation (22) while

\begin{equation}
\sigma=\frac{m(1-n)}{1+n}+1
\end{equation}

and
\begin{equation}
h=\frac{a}{\xi+\sigma}(-\nu_{_{-1}}-\sigma)\sqrt{F_{0}}\left(-\frac{F_{m}}
{2F_{0}}\right)^{\frac{1-n}{1+n}}
\end{equation}

We note that since here $m$ can have values 1 and 2 for $n=2$ and
$m=1$ for $n\ge 3$ so $\sigma$ is always positive. But for $h$ to
be positive definite, $\nu_{_{-1}}$ is restricted within the
range $-1\le \nu_{_{-1}}\le -\sigma$ i.e., $\nu_{_{-1}}$ is
negative definite.\\

(ii) for $\xi=m$, $a$ and $\xi$ have the same expressions as in
equation (23) and $\sigma$ is obtained from the equation

\begin{equation}
m+\sigma=2^{\frac{n-1}{n+1}}~b(n+1)\left[-\zeta-(n+1)b\right]^{-\frac{2n}{n+1}}
-2^{\frac{n-1}{n+1}}~\nu_{_{-1}}~F_{0}^{-1/2}\left[-\zeta-(n+1)b\right]^{\frac{1-n}{1+n}}
\end{equation}

with expressions for $b$ and $\zeta$ from equation (25), while the
other constant $h$ is totally arbitrary. Thus it is possible to
have an entire family of outgoing null geodesics terminated in
the past at the singularity, provided $\sigma$ obtained from
equation (31) is positive definite.\\

\section{{\protect\normalsize \textbf{Radial Time-like Geodesics}}}

We shall now investigate whether it is possible to have any
outgoing time-like geodesic originated from the singularity. For
simplicity of calculation we shall consider only outgoing radial
time-like geodesic (ORTG). Let us denote by
$K^{a}=\frac{dx^{a}}{d\tau}$, a unit tangent vector field to a
ORTG with $\tau$ an affine parameter along the geodesic. Hence
from the geodesic equation we have [20]

\begin{equation}
K^{t}\left(K^{r}\right)^{.}+2\dot{\alpha}K^{t}K^{r}+K^{r}\left(K^{r}\right)'+
\alpha'\left(K^{r}\right)^{2}=0
\end{equation}

with
\begin{equation}
K^{t}=\pm\sqrt{1+e^{2\alpha}\left(K^{r}\right)^{2}}~~.
\end{equation}

In order to satisfy the above two equations the simplest choice
for ($K^{r}, K^{t}$) is

\begin{equation}
K^{t}=\pm 1,~~~~K^{r}=0~.
\end{equation}

and so we have the solution

\begin{equation}
t-t_{0}=\pm(\tau-\tau_{0}),~~~~r=r_{0}=\text{constant}.
\end{equation}

Here $+$ (or $-$) sign corresponds to ORTG (or ingoing RTG) and
$\tau_{0}$ is the proper time at which radial time-like geodesic
passes through the central singularity.\\

Now similar to null geodesic let us choose the radial time-like
geodesic near the singularity to be of the form (to leading order)

\begin{equation}
t_{ORTG}(r)=t_{0}+c~r^{p}
\end{equation}

with $c$ and $p$ as positive constants. Further, consistent with
equations (34) and (35) we assume

\begin{equation}
K^{r}(t,r)=A(t-t_{0})^{\lambda}r^{\delta}
\end{equation}

where $A~(>0), \lambda$ and $\delta$ are constants. Hence we have

\begin{equation}
K^{r}(t_{ORTG},r)=A~r^{q},~~~~q=\lambda p+\delta.
\end{equation}

Also from the geodesic equation (36) using (38) we get

\begin{equation}
\frac{dt_{ORTG}}{dr}=cpr^{p-1}=\frac{K^{t}}{K^{r}}=\sqrt{A^{-2}r^{-2q}+(R'+R\nu')^{2}}
\end{equation}

Using the solution for $R$ near the singularity in equation (39)
and equating equal powers of $r$ we have\\

(i) $p=1-q$,~~$c=\frac{1}{A(1-q)}$~~ if~~
$-\frac{2m}{n+1}<q<1$.\\

(ii)
$p=1+\frac{2m}{n+1}$,~~$c=\frac{(\nu_{_{-1}}+1+\frac{2m}{n+1})}{(1+\frac{2m}{n+1})}
\left(-\frac{F_{m}}{2F_{0}}\right)^{\frac{2}{n+1}}$~~ if~~
$q<-\frac{2m}{n+1}$\\

(iii)
$p=1+\frac{2m}{n+1}$,~~$c=\frac{1}{p}\left[A^{-2}+\left(-\frac{F_{m}}{2F_{0}}
\right)^{\frac{4}{n+1}}\left(\nu_{_{-1}}+1+\frac{2m}{n+1}\right)^{2}\right]^{1/2}$~~
if~~ $q=-\frac{2m}{n+1}$\\

Therefore equation (36) has consistent solution for $c$ and $p$
depending on the parameters involved. So it is possible to have
outgoing radial time-like geodesic originated from the
singularity.

\section{{\protect\normalsize \textbf{Local Visibility}}}

We have shown the existence of outgoing both null and time-like
geodesics which were originated in the past from the singularity.
We shall now examine whether these geodesics are visible to any
non-space like observer. Now for visibility of the singularity
(local nakedness) any future directed geodesics (null or
time-like) starting from the singularity should be outside the
domain of dependence of any trapped surfaces both before and at
the time of formation of apparent horizon (AH) which is the outer
boundary of the trapped surface. Now if $t_{ah}(r)$ is the time
of formation of apparent horizon so we must have [8,9,18]
$$
\dot{R}(t_{ah}(r),r)=-1
$$
i.e.,
\begin{equation}
R^{n-1}(t_{ah}(r),r)=F(r)
\end{equation}

The apparent horizon is given by the curve

\begin{equation}
t_{ah}(r)=t_{0}-\frac{1}{(n+1)F_{0}^{3/2}}\left(F_{m}r^{m}+O(r^{m+1})\right)
-\frac{2}{n+1}F_{0}^{\frac{1}{n-1}}\left(r^{\frac{n+1}{n-1}}+O(r^{\frac{n+1}{n-1}+1})\right)
\end{equation}

As the apparent horizon and the singularity curve form at the
same time at $r=0$, so the visibility of the singularity is
determined by the relative slopes of the curve $t_{ah}(r)$ and
the curve for outgoing radial non-space like geodesics (denoted
by $t_{ORG}(r)$). Hence the necessary and sufficient condition
for singularity to be at least locally naked is that

\begin{equation}
\begin{array}{c}
lim\\
r\rightarrow 0\\
\end{array}
\begin{array}{c}
\left\{(\frac{dt_{ah}}{dr})/(\frac{dt_{ORG}}{dr})\right\}>1\\
{}
\end{array}
\end{equation}

Now we shall examine this condition for both ORNG and ORTG
separately.

\subsection{Outgoing Radial Null Geodesic}

In this case the ratio of the slopes is

\begin{eqnarray*}
\frac{(\frac{dt_{ah}(r)}{dr})}{(\frac{dt_{ORNG}(r)}{dr})}=-\frac{1}{a\xi(n+1)F_{0}^{3/2}}
\left(mF_{m}r^{m-\xi}+O(r^{m-\xi+1})\right)
\end{eqnarray*}
\vspace{-5mm}

\begin{equation}
~~~~~~~~~~~~~~~~-\frac{2}{a\xi(n-1)}F_{0}^{\frac{1}{n-1}}
\left(r^{\frac{n+1}{n-1}-\xi}+O(r^{\frac{n+1}{n-1}-\xi+1})\right)
\end{equation}

So we have the following possibilities:\\

(i) If $m>\frac{n+1}{n-1}$ then the above ratio of the slope is
always negative, hence the condition is violated. So we always
get black hole.\\

(ii) If $m<\frac{n+1}{n-1}$ then the above ratio of the slopes
approaches to $+\infty$ as $r\rightarrow 0^{+}$ for $m<\xi$ and
while the limit will be finite and greater than unity for $m=\xi$
provided $a<-\frac{m}{(n+1)\xi}\frac{F_{m}}{F_{0}^{3/2}}$. Thus
the singularity will be locally naked for any dimension if $m=1$
and $a<-\frac{1}{(n+1)\xi}\frac{F_{1}}{F_{0}^{3/2}}$ while naked
singularity appears only for four dimension if $m=2$ and
$a<-\frac{2}{3\xi}\frac{F_{2}}{F_{0}^{3/2}}$.\\

(iii) If $m=\frac{n+1}{n-1}$ then we have shown that naked
singularity appears only upto five dimension and the singularity
will be locally visible for
\[
F_{m}<-2F_{0}^{\frac{m}{2}+1},~~~~m<\xi
\]
\[\text{or}\]
\[
F_{m}<-2F_{0}^{\frac{m}{2}+1}-a(n+1)F_{0}^{3/2},~~~~m=\xi.
\]

\subsection{Outgoing Radial Time-like Geodesic}

In this case the ratio of the slopes is

\begin{eqnarray*}
\frac{(\frac{dt_{ah}(r)}{dr})}{(\frac{dt_{ORTG}(r)}{dr})}=-\frac{1}{cp(n+1)F_{0}^{3/2}}
\left(mF_{m}r^{m-p}+O(r^{m-p+1})\right)
\end{eqnarray*}
\vspace{-5mm}

\begin{equation}
~~~~~~~~~~~~~~~~-\frac{2}{cp(n-1)}F_{0}^{\frac{1}{n-1}}
\left(r^{\frac{n+1}{n-1}-p}+O(r^{\frac{n+1}{n-1}-p+1})\right)
\end{equation}

We shall now study the following possibilities for local
visibility:\\

(i) If $m>\frac{n+1}{n-1}$ then as before only black hole
appears.\\

(ii) If $m<\frac{n+1}{n-1}$ we have the conclusion as in ORNG
except that instead of restricting $a$, we have here
$c<-\frac{m}{(n+1)p}\frac{F_{m}}{F_{0}^{3/2}}$.\\

(iii) Similar is the situation for $m=\frac{n+1}{n-1}$.\\

Hence the conditions for local visibility of naked singularity is
consistent for both time-like and null geodesic.\\

\section{{\protect\normalsize \textbf{Discussion and Concluding Remarks}}}
In this paper we have studied quasi-spherical gravitational
collapse in $(n+2)$-D Szekeres space-time by studying both the
null and time-like geodesics originated from the central shell
focusing singularity. Following the approach of Barve et al [23]
for null geodesic we have shown that if we choose the parameter
$\nu_{_{-1}}$ to be greater than $-1$, then it is possible to
have a class of outgoing radial null geodesic for four and five
dimensions only with the restriction that initial density falls
off rapidly to the centre. However for $\nu_{_{-1}}=-1$, naked
singularity is possible in all dimensions irrespective of the
assumption on the initial density. On the other hand, for
time-like geodesic the results are very similar to the study of
null geodesic. Regarding local visibility we have consistent
results for both ORNG and ORTG with some restrictions on the
parameters involved in the equation of the geodesics. Therefore
we conclude that it is possible to have at least local naked
singularity for the given higher dimensional non-spherical
space-time.\\\\

{\bf Acknowledgement:}\\

One of the authors (U.D) thanks CSIR (Govt. of India) for the
award of a Senior Research Fellowship.\\

{\bf References:}\\
\\
$[1]$  A. Ori and T. Piran, {\it Phys. Rev. Lett.} {\bf 59} 2137 (1987);
J. P. S. Lemos {\it Phys. Rev. Lett.} {\bf 68} 1447 (1992).\\
$[2]$  P.S. Joshi, {\it Global Aspects in Gravitation and
Cosmology}(Oxford Univ. Press, Oxford, 1993).\\
$[3]$  P. S. Joshi and I. H. Dwivedi, {\it Commun. Math. Phys.}
{\bf 166} 117 (1994).\\
$[4]$  P. S. Joshi and I. H. Dwivedi, {\it Class. Quantum Grav.}
{\bf 16} 41 (1999).\\
$[5]$  F. J. Tipler, {\it Phys. Lett. A} {\bf 64} 8 (1987).\\
$[6]$  F. C. Mena, R. Tavakol and P. S. Joshi, {\it Phys. Rev. D}
{\bf 62} 044001 (2000).\\
$[7]$  P.S. Joshi, N. Dadhich and R. Maartens, {\it Phys. Rev. D}
{\bf 65} 101501({\it R})(2002).\\
$[8]$ A. Banerjee, U. Debnath and S. Chakraborty, {\it
gr-qc}/0211099 (2002)(accepted in {\it Int. J. Mod. Phys. D}) .\\
$[9]$ U. Debnath and S. Chakraborty, {\it gr-qc}/0211102 (2002) .\\
$[10]$ R. Goswami and P.S. Joshi, {\it gr-qc}/02112097  (2002) .\\
$[11]$ K. S. Thorne, 1972 in {\it Magic Without Magic}: John
Archibald Wheeler, Ed. Klauder J (San Francisco: W. H. Freeman
and Co.).\\
$[12]$  S. L. Shapiro and S. A. Teukolsky, {\it Phys. Rev. Lett.} {\bf 66} 994 (1991).\\
$[13]$ T. Nakamura, M. Shibata and K.I. Nakao, {\it Prog. Theor.
Phys.} {\bf 89} 821 (1993) .\\
$[14]$ T. Harada, H. Iguchi and K.I. Nakao, {\it Phys. Rev. D} {\bf 58} 041502 (1998) .\\
$[15]$  H. Iguchi, T. Harada and K.I. Nakao, {\it Prog. Theor.
Phys.} {\bf 101} 1235 (1999); {\it Prog. Theor. Phys.} {\bf 103} 53 (2000).\\
$[16]$ C. Barrabes, W. Israel and P. S. Letelier, {\it Phys. Lett.
A} {\bf 160} 41 (1991); M. A. Pelath, K. P. Tod and R. M.
Wald, {\it Class. Quantum Grav.} {\bf 15} 3917 (1998).\\
$[17]$  P. Szekeres, {\it Commun. Math. Phys.} {\bf 41} 55 (1975).\\
$[18]$  P. Szekeres, {\it Phys. Rev. D} {\bf 12} 2941 (1975).\\
$[19]$  P. S. Joshi and A. Krolak, {\it Class. Quantum Grav.}
{\bf 13} 3069 (1996); S. S. Deshingkar, S. Jhingan and P. S.
Joshi, {\it Gen. Rel. Grav.} {\bf 30} 1477 (1998).\\
$[20]$  S. M. C. V. Goncalves, {\it Class. Quantum Grav.}
{\bf 18} 4517 (2001).\\
$[21]$  S. Chakraborty and U. Debnath , {\it gr-qc}/0304072 (2003).\\
$[22]$  U. Debnath, S. Chakraborty and J. D. Barrow, {\it gr-qc}/0305075 (2003).\\
$[23]$  S. Barve, T. P. Singh, C. Vaz and L. Witten, {\it Class.
Quantum Grav.} {\bf 16} 1727 (1999).\\

\end{document}